\documentclass[epj,nopacs]{svjour}

\usepackage{graphicx}

\newcommand{\half}{{\textstyle{\frac12}}}
\newcommand{\thalf}{{\textstyle{\frac32}}}

\def\vec#1{\mbox{\boldmath$#1$}}

\begin{document}

\title{The $\Lambda(1405)$ resonance as a genuine three-quark 
or molecular state}

\author{%
B.~Golli \inst{1}\thanks{\email{bojan.golli@ijs.si}}
\and
H. Osmanovi\'c \inst{2}\thanks{\email{hedim.osmanovic@untz.ba}} 
\and
S.~\v{S}irca \inst{3}\thanks{\email{simon.sirca@fmf.uni-lj.si}}
}

\institute{%
Faculty of Education,
              University of Ljubljana and J.~Stefan Institute,
              1000 Ljubljana, Slovenia
\and
Faculty of Natural Sciences and Mathematics, University of Tuzla
              75000 Tuzla, Bosnia and Hercegovina
\and
Faculty of Mathematics and Physics,
              University of Ljubljana and J.~Stefan Institute,
              1000 Ljubljana, Slovenia}

\date{\today}

\abstract{%
The mechanism for the formation of the $\Lambda(1405)$ resonance
is studied in a chiral quark model that includes
quark-meson as well as contact (four point) interactions.
The negative-parity $S$-wave scattering amplitudes for strangeness 
$-1$ and 1 are calculated within a unified coupled-channel 
framework that includes the   $KN$,  $\bar{K}N$, $\pi\Sigma$, 
$\eta\Lambda$, $K\Xi$, $\pi\Lambda$, and $\eta\Sigma$ channels 
and possible genuine three-quark bare singlet and octet states
corresponding to $\frac12^-$ resonances.
It is found that in order to reproduce the scattering amplitudes in
the $S_{01}$ partial wave it is important to include the 
pertinent three-quark octet states as well as the singlet state,
while the inclusion of the contact term is not mandatory.
The  Laurent-Pietarinen expansion is used to determine 
the $S$-matrix poles.
Following their evolution as a function of increasing interaction 
strength, the mass of the singlet state is strongly reduced due to 
the attractive self-energy in the $\pi\Sigma$ and $\bar{K}N$ channels; 
when it drops below the $KN$ threshold, the state acquires a dominant
$\bar{K}N$ component which can be identified with a molecular state.
The attraction between the kaon and the nucleon is generated through
the $\bar{K}N\Lambda^*$ interaction rather than by meson-nucleon forces.}

%\pacs{{}11.80.Gw, 12.39.Ba, 14.20.Gk}

%\keywords{chiral quark models, baryon resonances}

\maketitle

\section{Introduction}
The lowest state with negative parity in the strange sector, the 
$\Lambda(1405)$ resonance, has been receiving particular attention 
since its discovery six decades ago~\cite{Alston61}.
The intriguing property of this resonance is that it lies
some 100~MeV below the corresponding negative parity state in 
the nonstrange sector, the $N(1535)$, which cannot be explained
in the ordinary quark model involving only three quarks.

The quark model calculations in the $S_{01}$ partial wave, 
nonrelativistic or relativistic
\cite{Isgur78,Isgur85,Isgur86,Metsch01,Plessas08},
assuming that the three lowest negative parity and strangeness 
$-1$ resonances correspond to the three-quark states in which 
one quark is excited to the $p$ orbit, indeed predict the mass 
of the lowest state to be around 1500~MeV, while the masses of 
the other two states turn out to be consistent with the masses 
of the non-strange negative parity resonances.
To resolve the discrepancy between the quark model prediction 
of the $\Lambda(1405)$ mass and the observed one,  
Arima et al.~\cite{Arima94} included explicit $\bar{K}N$ 
and $\pi\Sigma$ configurations and showed that the significant 
downward shift of the $\Lambda(1405)$ mass can be attributed 
to the large attractive self-energy due to these additional 
degrees of freedom.

The idea that the $\Lambda(1405)$ is predominantly a $\bar{K}N$
bound system without explicit quark degrees of freedom has been 
pursued by several groups starting with the work  of Dalitz et 
al.~\cite{Dalitz60,Dalitz67} based on a vector exchange model.
Most of the calculations were performed in the coupled-channel 
framework in a chiral unitary approach~\cite{Weise95,Oset98,Lutz02}.
This approach leads to the two-pole picture of the resonance~\cite{%
Ulf01,Ulf03,Hosaka03,Arriola03,Hyodo08,Ikeda11,Ikeda12,Guo13,Roca13,Ulf15} 
with a narrow pole just below the $KN$ threshold, and a wider pole at 
the mass either around 1380~MeV, or close to the $\pi\Sigma$ threshold.
According to Myint et al.~\cite{Myint18} only the first pole 
produces a peak in the observed spectrum while the second pole affects 
only the shape of the detectable spectrum.

A model that is able to incorporate both the genuine (bare) 
three-quark states as well as the baryon-mesons pairs in a consistent 
approach is the Cloudy Bag Model~\cite{CBM} (CBM).
In the $SU(3)$ extended version of the CBM~\cite{Thomas84,Thomas85,Jennings86} 
with a bare three-quark singlet state representing the $\Lambda(1405)$ 
and the baryon-meson configurations corresponding to  $\bar{K}N$ and 
$\pi\Sigma$ channels, the authors were able to obtain a resonance below 
the  $KN$ threshold and dominated by a  $\bar{K}N$ molecular state.
The $KN$ system was studied in the same model in Ref.~\cite{Thomas85b}.

Lattice calculations~\cite{Leinweber12,Leinweber15},
interpreted in the framework of Hamiltonian effective theory
\cite{Leinweber18}, reveal a dominant $\bar{K}N$ component
in the  light quark-mass regime.
The Graz group~\cite{Lang13} confirmed a non-negligible
singlet three-quark component with an admixture
of the octet states at a level of 15~\% -- 20~\%.
The calculation of Gubler et al.~\cite{Gubler16} confirms the
dominant singlet component for the lowest lattice $\half^-$ state,
and identifies the second lattice state with the $\Lambda(1670)$.
Their calculation is analyzed using hadronic effective theory in 
a finite volume in Ref.~\cite{Gubler21}; 
for pion masses above 290~MeV the lowest lattice state is identified 
with only one of the two states predicted by the chiral unitary 
approach.

Partial wave analyses for $K^-p$ scattering have been performed
by the Kent group~\cite{Manley13a,Manley13b}, 
Kamano et al.~\cite{Kamano14} and the Bonn-Gatchina
group~\cite{Sarantsev19a,Sarantsev19b}.
Kamano et al.~\cite{Kamano15} predicted two poles below the 
$KN$ threshold, while the Bonn-Gatchina group~\cite{Anisovich20}
found only one physically convincing pole in this region.
Fern\'andez-Ram\'irez et al.~\cite{Fernandez16} have discovered 
that the pole near the $KN$ threshold belongs to the $0^-$ 
leading Regge trajectory and therefore is most likely dominated 
by the ordinary three-quark configuration.
Similarly Klempt et al.~\cite{Klempt20} conclude that one of 
the two states found as poles below the $KN$ threshold has to be 
assigned to the predicted quark-model state.

In order to investigate the dominant mechanism responsible for 
the resonance formation in the $S_{01}$ partial wave, we devise 
a model that incorporates dynamical generation 
as well as generation through a three-quark resonant state.
In a similar approach we have been able to show, for instance, 
that the Roper resonance evolves from a genuine three-quark 
state~\cite{PRC2018}, while its $I=J=\frac32$ partner, 
the $\Delta(1600)$, emerges as a purely dynamically generated 
resonance~\cite{PRC2019}. 

We use a coupled-channel formalism incorporating  qua\-si-bound 
quark-model states to calculate the meson-pro\-duction amplitudes
in the $S_{01}$ and $S_{11}$ partial waves.
The meson-baryon vertices and the contact (four point) interaction
are determined in a chiral quark model; in the present approach
we use an $SU(3)$ extended  version of the CBM~\cite{Thomas85}.
The method of including bare three-quark states in the 
coupled-channel formalism has been described in detail in our 
previous papers~\cite{%
PRC2018,PRC2019,EPJ2008,EPJ2009,EPJ2011,EPJ2013,EPJ2016}
where we have analyzed scattering and electro-production 
amplitudes in different partial waves in the non-strange sector.
In the present work we consider the $KN$ channel in the 
strangeness $S=1$ sector and for $S=-1$ the coupled channels
$\pi\Sigma$, $\bar{K}N$, $\eta\Lambda$, $K\Xi$, $\pi\Lambda$, and 
$\eta\Sigma$, as well as the bare three-quark states corresponding 
to the $\Lambda(1405)$, $\Lambda(1670)$, $\Lambda(1800)$, 
$\Sigma(1750)$, and $\Sigma(1900)$ resonances.

In the next section we briefly review the basic features of 
our coupled-channel approach and of the underlying quark model.
We write down the Lippmann-Schwinger equation (LSE) for the meson
amplitudes in the $K$-matrix approach and discuss the construction 
of its kernel and the related background (nonresonant) terms.
In Sec.~\ref{sec:scattering} we solve the coupled-channel 
system in the $S_{01}$ and $S_{11}$ partial waves, starting with 
the $S=1$ sector, which provides information of the relevant 
nonresonant processes, and continue with the $S=-1$ sector, 
first trying to obtain resonances without bare quark states, 
and then including the pertinent bare quark states. 
Our primary aim is not to fine-tune the model parameters in order 
to fit the data but rather to investigate whether the model 
parameters used in the nonstrange sector, as well as the couplings 
evaluated in the quark model, are  able to reproduce the main 
features of the experimental amplitudes.
In Sec.~\ref{sec:structure} we analyze the properties of the
resonances and their relation to the bare quark states by 
following the evolution of the poles in the complex energy plane. 
Special emphasis is given to the $\Lambda(1405)$ resonance, 
analyzing the interplay of the bare-quark and molecular degrees 
of freedom.

\section{\label{sec:model} The model}
\subsection{The coupled channel approach for the $K$ matrix}

In our approach the scattering state in channel $\alpha$ which includes  
quasi-bound quark states $\Phi^0_i$, $i=1,\ldots, N_r$ assumes the form
\begin{eqnarray}
   |\Psi_\alpha\rangle &=& \mathcal{N}_\alpha\biggl\{
        \left[a^\dagger_\alpha(k_\alpha)|\Phi_\alpha\rangle\right]
   +  \sum_{i=1}^{N_r} c_{\alpha i}|\Phi^0_i\rangle
\biggr.\nonumber\\ && \biggl. 
  +  \sum_\beta
   \int 
   {dk\>\chi_{\alpha\beta}(k_\alpha,k)\over\omega_\beta(k)+E_\beta(k)-W}\,
   \left[a^\dagger_\beta(k)|\Phi_\beta\rangle\right]\biggr\},
\label{PsiH}
\end{eqnarray}
where $\alpha$ ($\beta$) denote the channels and [ ] stands for 
coupling to the appropriate total spin and isospin.
The first term represents the free meson and the baryon and defines 
the channel, the second term corresponds to the sum over $N_r$ 
{\em bare\/} three-quark resonant states\footnote{In our previous 
calculations we have included only one or two quasi-bound quark 
state; in the present work we allow for more quark states.}, 
while the third term describes the meson cloud around the baryon 
in channel $\beta$.
All quantities are written in the center-of-mass frame:
$\omega_\alpha(k)$ and $E_\alpha(k)$ are, respectively, the meson 
and the baryon off-shell energies in channel $\alpha$, 
the on-shell values are denoted as $k_\alpha$, 
$\omega_\alpha\equiv\omega_\alpha(k_\alpha)$ and 
$E_\alpha\equiv E_\alpha(k_\alpha)$, $W=\omega_\alpha+E_\alpha$ 
is the invariant energy, and 
$\mathcal{N}_\alpha=\sqrt{\omega_\alpha E_\alpha/(k_\alpha W)}$
is the normalization factor.
The integral is assumed in the principal value sense.
The (half-on-shell) $K$ matrix is related to the scattering state 
by~\cite{EPJ2008}
\begin{equation}
   K_{\alpha\beta}(k_\alpha,k) = -\pi\mathcal{N}_\beta
  \langle \Psi_\alpha||V^\beta(k)||\Phi_\beta\rangle\,,
\label{eq4K}
\end{equation}   
with the property $K_{\alpha\beta}(k_\alpha,k) = K_{\beta\alpha}(k,k_\alpha)$.
The meson-baryon interaction $V^\beta$ in channel $\beta$ is
explicitly written out in Appendix~\ref{vertices}.
The $K$ matrix is proportional to the meson amplitude $\chi$ 
in Eq.~(\ref{PsiH}),
\begin{equation}
   K_{\alpha\beta}(k_\alpha,k)
       = \pi\,\mathcal{N}_\alpha\mathcal{N}_\beta\,
             \chi_{\alpha\beta}(k_\alpha,k) \,.
\label{chi2K}
\end{equation}
The principal-value states (\ref{PsiH}) are normalized as
\begin{equation}
   \langle\Psi_\alpha(W) | \Psi_\beta(W')\rangle
  = \delta(W-W') \left[\delta_{\alpha\beta} 
  + {K^2}_{\alpha\beta}\right]\,.
\label{normPV}
\end{equation}
They are not orthonormal; the orthonormalized states
are constructed by inverting the norm.

The amplitude $\chi$ satisfies a Lippmann-Schwinger type 
of equation:
\begin{eqnarray}
   \chi_{\alpha\gamma}(k,k_\gamma) 
   &=& -\sum_i{c}_{\gamma i}\, V_{\alpha i}(k)
       + \mathcal{K}_{\alpha\gamma}(k,k_\gamma)
\nonumber\\ && 
+ \sum_\beta\int dk'\;
  {\mathcal{K}_{\alpha\beta}(k,k')\chi_{\beta\gamma}(k',k_\gamma)
  \over \omega_\beta(k') + E_{\beta}(k')-W}\,.
\label{eq4chi}
\end{eqnarray}
The explicit expression for the kernel $\mathcal{K}_{\alpha\beta}$ 
will be discussed in the next subsection.

The meson amplitude can be written in terms of the resonant and
nonresonant parts,
\begin{equation}
   \chi_{\alpha\gamma}(k,k_\gamma) = 
    \sum_i c_{\gamma i}{\cal V}_{\alpha i}(k) +
    {\cal D}_{\alpha\gamma}(k,k_\gamma)\,,
\label{splitchi}
\end{equation}
such that~(\ref{eq4chi}) can be split into $N_r$ equations for the
dressed vertices,
\begin{equation}
\mathcal{V}_{\alpha i}(k)
= V_{\alpha i}(k)
 + \sum_{\beta}  \int dk'\;
        {\mathcal{K}_{\alpha\beta}(k,k')
         \mathcal{V}_{\beta i}(k')
   \over   \omega_\beta(k')+E_\beta(k')-W}\,,
\label{eq4VR}
\end{equation}
and an equation for the nonresonant amplitude,
\begin{equation}
\mathcal{D}_{\alpha\gamma}(k,k_\gamma) = 
\mathcal{K}_{\alpha\gamma}(k,k_\gamma)
+  \sum_\beta \int dk'\;
        {\mathcal{K}_{\alpha\beta}(k,k')
         \mathcal{D}_{\beta\gamma}(k',k_\gamma)
   \over   \omega_\beta(k')+E_\beta(k')-W}\,.
\label{eq4D}
\end{equation}

By requiring stationarity, 
$\langle\delta\Psi_\alpha|H-W|\Psi_\alpha\rangle =0$, 
with respect to variation of the coefficients $c_{\alpha i}$
we get a system of equations:
\begin{equation}
\sum_jA_{ij}c_{\alpha j} = \mathcal{V}_{\alpha i}
\end{equation}
where
\begin{equation}
A_{ij} = (W - m_i^0 )\delta_{ij}
    + \sum_\beta\int dk\;{\mathcal{V}_{\beta i}(k){V}_{\beta j}(k)
                \over \omega_\beta(k)+E_\beta(k)-W}\,.
\end{equation}
The matrix corresponding to this system is singular if 
${\rm det}{A}(W) = 0$; at those $W$, the coefficients $c_{\alpha i}$ 
and consequently the $K$ matrix have poles.
It is convenient to solve the system by first diagonalizing ${A}$: 
$$
    {U}^T{A}{U} 
   = \hbox{diag}[Z_1^{-1}(W-m_1), Z_2^{-1}(W-m_2), \cdots] \,,
$$
and then explicitly invert it:
$$
    \left(A^{-1}\right)_{ji} =\sum_r{U_{jr}U_{ri}\,Z_r\over W-m_r}
$$
The resulting resonant part of the $K$ matrix can be cast in
the form
$$
     K^{\rm res}_{\alpha\gamma} 
   = \pi\mathcal{N}_\alpha\mathcal{N}_\gamma \sum_r
   {\widehat{\mathcal{V}}_{\alpha r} \widehat{\mathcal{V}}_{\gamma r} 
      \over m_r - W}\,,
\quad
   \widehat{\mathcal{V}}_{\alpha r} 
  = \sqrt{Z}_r\sum_i U_{ri}\mathcal{V}_{\alpha i}\,.
$$
Here $\sqrt{Z}_r$ is the wave-function renormalization while $U_{ri}$
are expansion coefficients of the physical resonance $r$ in terms
of the bare three-quark states.

The $T$ matrix is finally obtained by solving the Heitler equation 
$T=K + \mathrm{i}KT$.

\subsection{\label{CBM} The underlying quark model}

The vertices are calculated in a version of the Cloudy Bag Model
extended to the pseudo-scalar $SU(3)$ meson octet \cite{Thomas85}.
Since we study here only the $S_{01}$ and $S_{11}$ partial-wave 
resonances, only the  $s$-wave mesons are included, while for 
the resonant states it is assumed that one of the three quarks 
is excited from the $1s$ state to the $1p_{1/2}$ state. 

The interaction part of the Hamiltonian consists of the
quark-meson part and the contact interaction:
\begin{eqnarray}
    \hat{H}_I &=& \hat{H}_s + \hat{H}_c \,,
\nonumber \\
    \hat{H}_s &=& \sum_\alpha\int dk[\hat{V}^\alpha(k) a_\alpha(k) + h.c.]\,,
\label{qM}\\
    \hat{H}_{c,\alpha\beta} &=& \int dk\,dk'\,\hat{V}^c_{\alpha\beta}(k,k') 
             a_\alpha^\dagger(k)a_\beta(k')\,.
\label{contact}
\end{eqnarray}
Here $\hat{V}^\alpha(k)$ is the quark operator of 
the surface (volume) part of interaction, while $a_\alpha$ is
the meson annihilation operator in channel $\alpha$.
The explicit expressions for the quark operators related
to pions, kaons and eta mesons are given in the Appendix.

The parameters of the model include the bag radius $R$ and the meson 
decay constants.
We use $R = 0.83$~fm and  $f_\pi=f_\eta=f_K=73$~MeV.
The latter value, smaller than the experimental one, is consistent 
with the value used in the ground-state calculations;
in particular it reproduces the $\pi NN$ coupling constant.
In the present calculation these two values have been fixed
in the meson-quark interaction.
On the other hand, since such a small value of $f_\pi$ may 
too strongly enhance the strength of the contact term, 
its value in the contact term will be considered as a free parameter.

In addition, the bare masses of the resonances are also free parameters, 
while the meson masses ($\mu_\alpha$) and the baryon masses ($m_\alpha$) 
in channel $\alpha$ are kept at the experimental values.

\subsection{The kernel of the LS equation}

Our previous calculation in the non-strange sector has
shown that the background can be well described through
the $u$-channel exchange processes alone.
For $S$-partial waves, the contact interaction may also play 
an important role.

For the kernel of the LSE we assume, apart from the contact
term, a term that reduces to the $u$-channel exchange term
when evaluated (half) on-shell:
\begin{eqnarray}
{\mathcal{K}}_{\alpha\beta}(W,k,k')  
&=& {V}^c_{\alpha\beta}(k,k') 
   + \sum_if^i_{\alpha\beta}\;V^\alpha_{i\beta}(k)\;V^\beta_{i\alpha}(k')
\nonumber\\
&&\times
\left[{1\over \omega_\beta(k') + \omega_\alpha(k) + E_i - W}\right.
\nonumber\\
&&
\left.  -{1\over \omega_\beta(k') + \omega_\alpha(k) - E_i - W}
\right]\,.
\label{kernel-ns}
\end{eqnarray}
Here $E_i\equiv E_i(\vec{k}+\vec{k'})
\approx\sqrt{m^2_i+k^2+{k'}^2}$ is the energy of the exchange
baryon for which  the resonances ($X^*$ and $S^*$ listed in 
table~\ref{tab:gus} and \ref{tab:gun}) may be considered.
For the $s$-wave mesons only the isospin quantum numbers of baryons 
($I_{\alpha(\beta)}, I_i$) and meson ($t_{\alpha(\beta)}$) are involved:
\begin{eqnarray}
f^i_{\alpha\beta} = \sqrt{(2I_\alpha+1)(2I_\beta+1)}
                 W(t_\alpha I_\beta I_\alpha t_\beta; I_i I)\,,
\end{eqnarray}
where $W(\cdots)$ are the Racah coefficients.
In our previous calculations we have used a separable approximation
for the $u$-exchange potential of the kernel (\ref{kernel-ns}):
\begin{eqnarray}
{\mathcal{K}}^{\rm sep}_{\alpha\beta}(W,k,k')  
&=&
     \sum_if^i_{\alpha\beta} {m_i\over E_{\alpha}}\,
  (\omega_{\alpha}+\varepsilon_{i\beta}^\alpha)
\nonumber\\ &\times&
   {V^\alpha_{i\beta}(k)\;V^\beta_{i\alpha}(k')
       \over (\omega_\alpha(k)+\varepsilon_{i\beta}^\alpha)
       (\omega_\beta(k')+\varepsilon_{i\alpha}^\beta)}\,,
  \label{kernel-sep}
\\
\varepsilon^\beta_{i\alpha} 
    &=& {m_i^2-m_\alpha^2 - \mu_\beta^2\over 2 E_\alpha}\,.
\end{eqnarray}
Here $\omega_{\beta}$, $\omega_{\alpha}$, $E_{\alpha}$, 
$\varepsilon_{i\beta}^\alpha$,  $\varepsilon_{i\alpha}^\beta$
are evaluated on-shell.
When one of the mesons is on-shell, both forms reduce to the
same expression.
In the present approach all baryons appearing in the channels are 
stable and the LSE can be solved numerically rather easily; since 
in this partial wave the meson-baryon couplings are relatively small, 
both full (unseparable) and separable form yield  very similar results.
However, since the contact term cannot be written in a separable form,
the LSE has to be anyway solved numerically.

\section{\label{sec:scattering} Solving the scattering equation}
\subsection{Solution for $KN$ scattering}

For strangeness $S=1$ there are no baryon resonances
and only background processes govern $KN$ scattering.
Furthermore, for isospin $I=0$ there is no contribution from
the contact term. 
This gives us an opportunity  to examine the validity of our 
approximation for the background term which stems from 
the $u$-channel exchange potential.
The scattering amplitudes are obtained by solving Eq.~(\ref{eq4D}).

It turns out that the main contribution comes from the exchange 
of the singlet $\Lambda^*_1$ baryon which can be identified with 
the $\Lambda(1405)$ resonance and the octet ${}^2\Lambda^*_8$ baryon 
identified with the $\Lambda(1800)$ resonance.
The identification of the octet and decouplet $\Sigma^*$ is
less clear and we do not consider their contribution here.

\begin{figure}[h]
\begin{center}
\includegraphics[width=75mm]{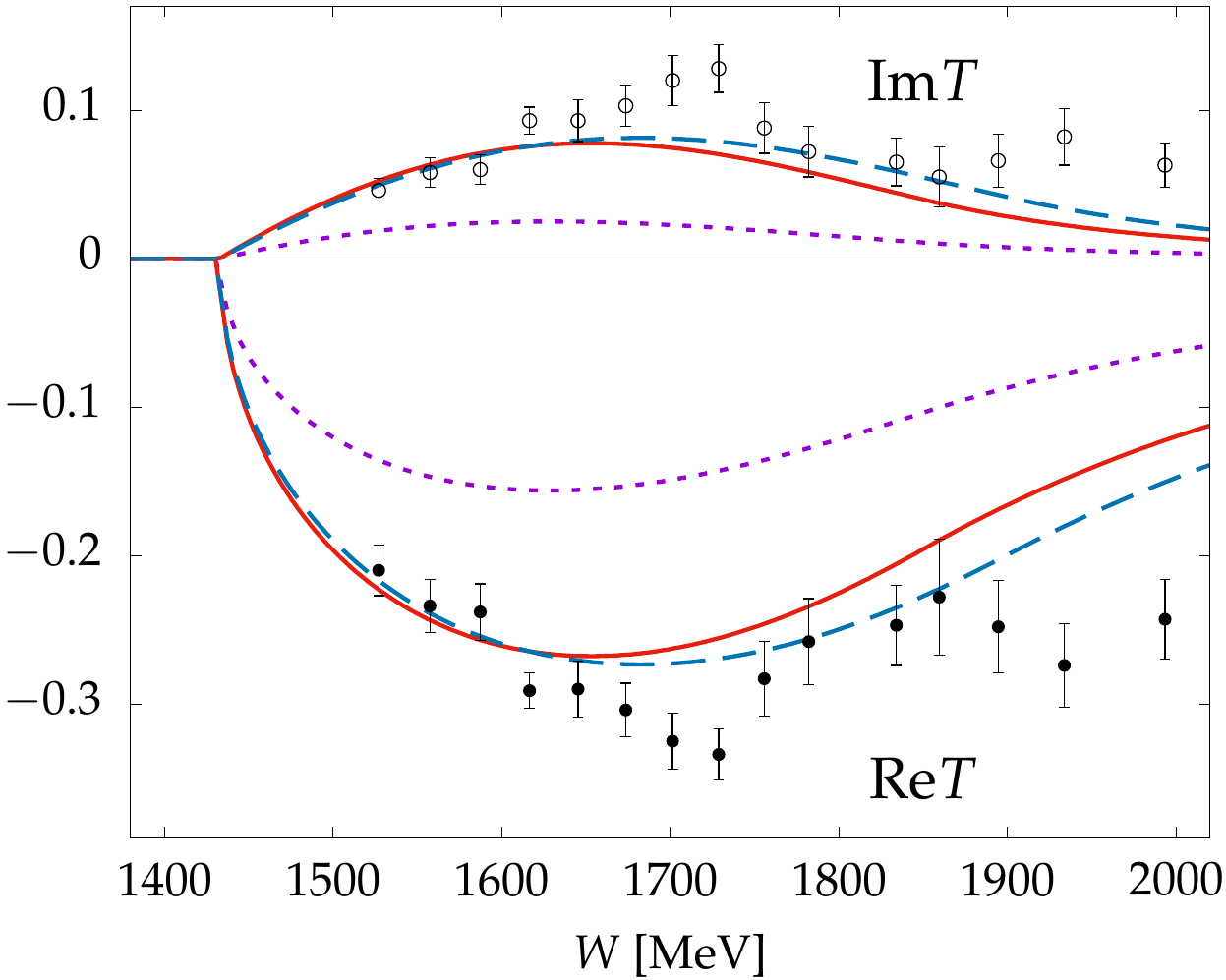}
\end{center}
\vspace{-12pt}
\caption{$T$-matrix amplitudes for the reaction $KN\to KN$ in the 
$S_{01}$ partial wave for different bag radii and interaction strengths.
(See text for curve assignments.) Experimental data are from~\cite{SAID}.} 
\label{fig:reimTKNI0}
\end{figure}

Using our choice of $R=0.83$~fm and $f_\pi=73$~MeV, as well as 
the coupling constants predicted by the quark model 
(table~\ref{tab:gus}), we underestimate the experimental 
$KN$ amplitudes (fig.~\ref{fig:reimTKNI0}, short dashes). 
Multiplying the coupling constants by a renormalization 
factor $f_u =1.35$ we are able to reproduce the amplitudes 
at low and intermediate energies (solid line).
The agreement is improved if we choose a slightly smaller
bag radius, $R=0.78$~fm, and $f_u =1.30$ (dashes).

In the following calculation we use the form and the renormalization 
factor derived above; we retain our standard choice of the bag radius, 
$R=0.83$~fm, which, as we see in the following, yields the most
consistent results in other sectors.

\begin{figure}[h]
\begin{center}
\includegraphics[width=75mm]{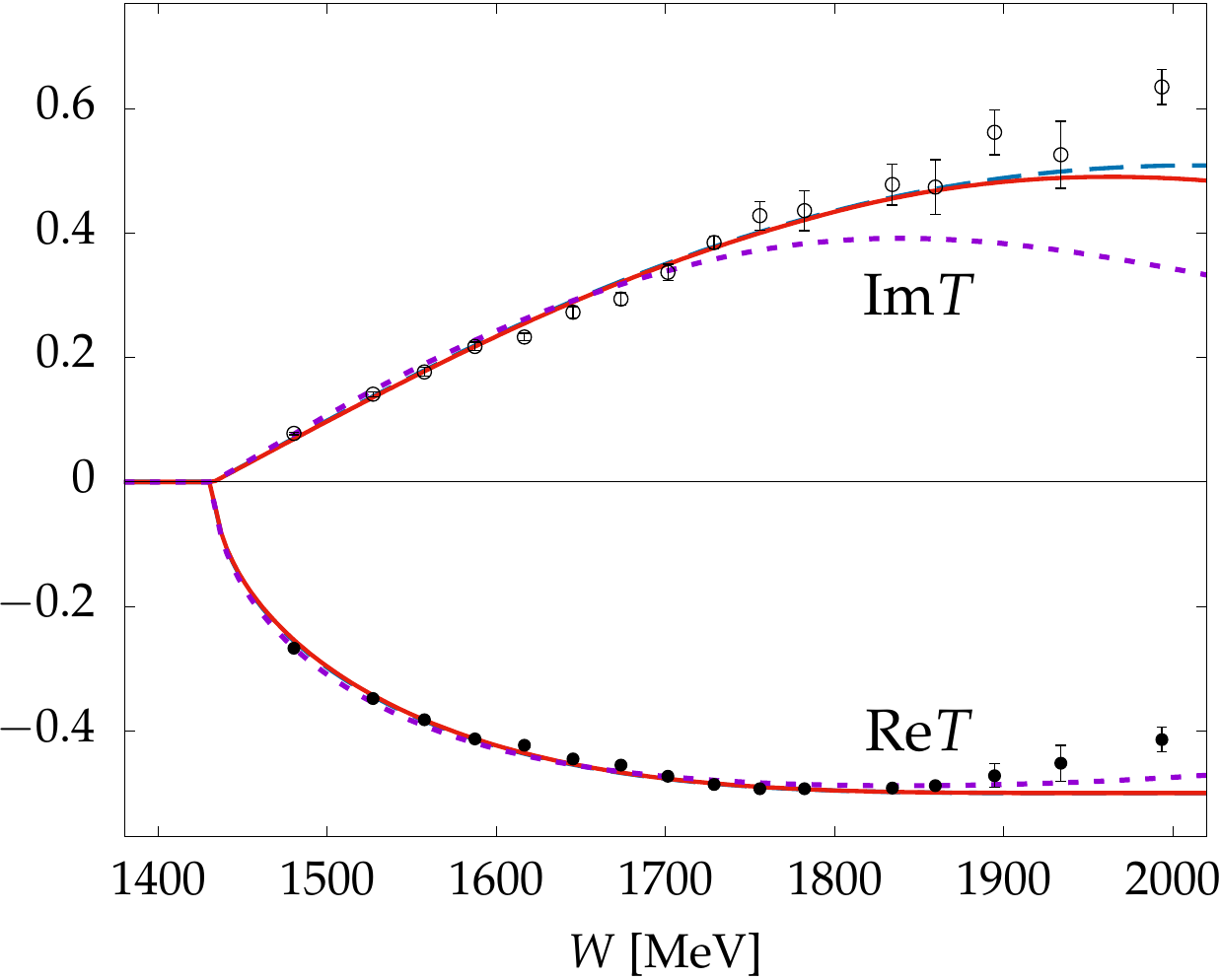}
\end{center}
\vspace{-12pt}
\caption{Same as fig.~\ref{fig:reimTKNI0}, 
but for the $S_{11}$ partial wave.} 
\label{fig:reimTKNI1}
\end{figure}

In the isospin $I=1$ channel (fig.~\ref{fig:reimTKNI1}) the contact 
interaction is present and dominates the amplitude; the exchange 
potential is relatively small and has the opposite sign with respect 
to the $I=0$ case as well as with respect to the contact potential.
We obtain a good agreement with the contact potential alone using 
the rather standard choice of $R=0.83$~fm and the experimental
value $f_\pi=93$~MeV (dashes); 
adding the exchange potential the experimental data are equally
well reproduced by taking $R=0.90$~fm and $f_\pi=85$~MeV (solid line).
Similarly as in the $I=0$ case the data support bag radii close
to those used to describe baryons.
Larger radii, e.g. $R=1$~fm (short dashes), underestimate
the Im$T$ part of the amplitude at higher $W$.

\subsection{Dynamically generated  $\Lambda(1405)$}

\begin{figure}[h]
\begin{center}
\includegraphics[width=70mm]{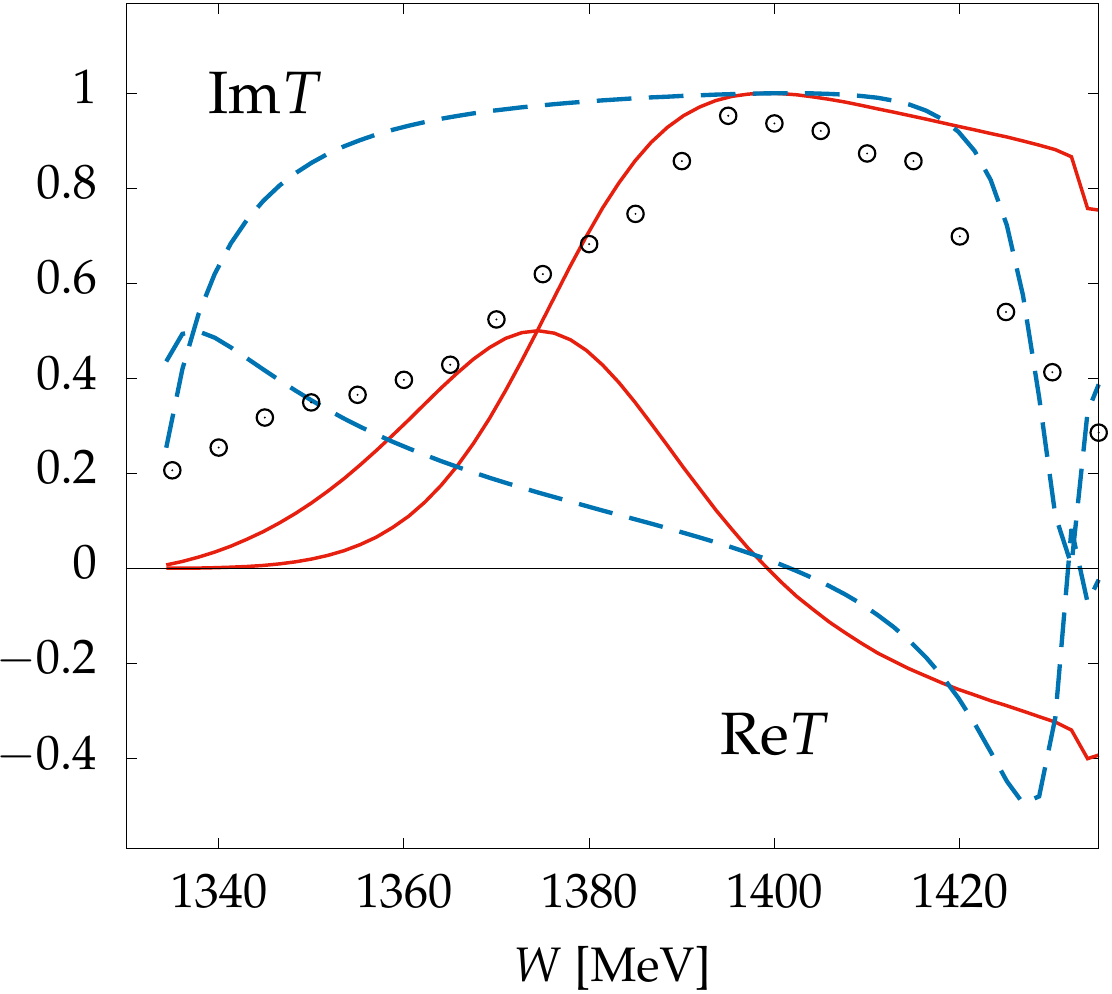}
\end{center}
\vspace{-12pt}
\caption{The real and imaginary part of the $\pi\Sigma\to\pi\Sigma$ 
amplitude  for $R=0.83$~fm and $f_\pi=78$~MeV (solid line), 
and for $R=1.1$~fm and $f_\pi=93$~MeV (dashes).
Since below the $KN$ threshold only the $\pi\Sigma$ channel is open, 
Im$T$ is proportional to the invariant mass distribution of 
$\pi\Sigma$ pairs and is compared to the experimental points 
in \cite{Hemingway1985}.}
\label{fig:TpiSdyn}
\end{figure}

Switching to the strangeness $S=-1$ sector,
we first consider the case with no bare three-quark states.
We solve the LSE (\ref{eq4D}) by using two channels, $\pi\Sigma$
and $\bar{K}N$, and assuming only contact interaction.
We first use our standard choice of model parameters,
$R=0.83$~fm and $f_\pi=73$~MeV.
The position of the pole in the complex $W$ plane
is calculated by using the Laurent-Pietarinen 
expansion~\cite{L+P2013,L+P2014,L+P2015}.
We obtain two poles: the upper pole very close to the $KN$ threshold 
and the lower one approaching the $\pi\Sigma$ threshold
(table~\ref{tab:poles-dyn}).
Reducing the strength of the interaction by assuming slightly
larger values for $f_\pi$, the mass of the lower pole rises to 
the nominal value given in \cite{PDG2020}, while the upper pole 
remains close to the $KN$ threshold.
The width of the upper pole is consistent with the PDG value
while the width of the lower pole seems to be underestimated by
a factor of two.
Let us note that in this case there is only one pole of the
$K$ matrix, which lies close to its nominal Breit-Wigner mass 
(fig.~\ref{fig:TpiSdyn}).

If we further reduce the strength of the contact term, only one 
pole remains, with a mass slightly below the $KN$ threshold.
This pole corresponds to the pole found in the same model in 
Ref.~\cite{Thomas85}.

\begin{table}[h]
\caption{\label{tab:poles-dyn}
The pole parameters obtained from the $S_{01}$ partial wave amplitudes 
in the $\pi\Sigma$ channel using the Laurent-Pietarinen expansion}
{\renewcommand{\arraystretch}{1.3}
\begin{center}
\setlength{\tabcolsep}{3pt}       
\begin{tabular}{lccccr} 
\hline
resonance & ~~Re$W$ & $-2{\rm Im}W$ & Module & $R$ & $f_\pi$ \\
     & [MeV] & [MeV]  & [MeV]   &  [fm] & [MeV] \\ 
\hline
$\Lambda(1380)$ & 1348 & 33 & 16 & 0.83 &  73  \\ 
                & 1378 & 48 & 20 & 0.83 &  78  \\ 
\hline
$\Lambda(1405)$ & 1433 & 20 &  1 & 0.83 &  73  \\ 
                & 1435 & 18 &  1 & 0.83 &  78  \\ 
                & 1430 & 10 &  3 & 1.00 & 100  \\
                & 1428 & 14 &  5 & 1.10 &  93  \\
\hline
\end{tabular}
\end{center}} 
\end{table}

Our results with smaller values of $R$ and $f_\pi$
are consistent with the predictions of the chiral unitary theory
and seem to suggest that no bare three-quark states 
are needed to reproduce at least the lowest resonance in 
the $S_{01}$ partial wave.
There is a caveat, however:
if we want to reproduce the subsequent two $I=0$ resonances, 
the $\Lambda(1670)$ and $\Lambda(1800)$, we have to assume 
the existence of at least two genuine quark model states.
Yet carrying out such a calculation we find that the 
required strength of the contact potential, necessary to support 
dynamically generated resonances below the $KN$ threshold, is 
much too large in order to reproduce the experimental scattering 
amplitudes in the region of the upper two $S_{01}$ resonances.
In fact, Kamano et al.~\cite{Kamano14}, who have done a
rather extensive analysis of the partial-wave amplitudes for
$K^-p$ scattering, have found a better agreement for the 
$S_{01}$ case {\em without\/} including the contact interaction.

\subsection{Including bare three-quark states in the $S_{01}$ partial wave}

For the $S_{01}$ partial wave we include the quark model states 
corresponding to the lowest three resonances, assuming one quark 
is promoted from the $s$ orbit to the $p_{1/2}$ orbit.
We further assume one singlet configuration, that can be identified
as the $\Lambda(1405)$, and two octet configurations, with internal
spin $S={1\over2}$ (doublet) and  $S={3\over2}$ (quadruplet)
that can be identified as the $\Lambda(1670)$ and $\Lambda(1800)$.
We use the $j$-$j$ coupling scheme identical to the one
used for the non-strange $S_{11}$ resonances~\cite{EPJ2011}.
The bare mass of the singlet state has been fixed by requiring that
a pole of the $K$ matrix lies at $W=1405$~MeV; the masses and 
a possible mixing angle of the bare octet states are free parameters.

We consider four channels: $\pi\Sigma$, $\bar{K}N$, $\eta\Lambda$, 
and $K\Xi$, and assume that the physical $\eta(548)$ 
implies $\eta=\eta_8\cos\theta_P-\eta_1\sin\theta_P$ with 
$\theta_P=-11.3^\circ$~\cite{PDG2020}.
We are not interested in obtaining the best fit to the experimental 
amplitudes but rather to investigate to what extent the quark model 
is able to reproduce the main features of the scattering amplitudes.
We therefore retain the quark-model values in table~\ref{tab:gs} 
for the first two channels, as well as for $\eta_8\Lambda\Lambda^*_1$.
The measured cross-section for $K^-+p\to\eta\Lambda$ \cite{BNL}
imposes a rather strong constraint on $\eta\Lambda$ coupling to the
octet $\Lambda^*_8$ and suggests a much smaller value for this 
coupling than the one predicted by the quark model;
similarly we take smaller values for $g_{K\Xi\Lambda^*}$.
We further assume our standard choice for the bag radius of
$R=0.83$~fm for all pertinent baryons as well as for the
decay constants $f_\pi=f_\eta=f_K=73$~MeV.

The background potential (and the kernel entering the LSE) consists 
of the $u$-channel exchange potential and the contact potential. 
Based on our discussion of $KN$ scattering for $I=0$ we keep
beside the nonstrange baryons in table~\ref{tab:gun}
only the $\Lambda^*$ as the exchange baryons with $S=-1$, 
and further assume a renormalization of the coupling constants 
in table~\ref{tab:gus} by a factor of $f_u=1.35$  for the $\bar{K}N$ 
as well as for the $\pi\Sigma$ channels.
We control the strength of the contact term by adjusting the 
value of $f_\pi$ which is allowed to differ from the (fixed) value
used in the meson-baryon interaction.

\begin{figure*}[t]
\begin{center}
\includegraphics[width=170mm]{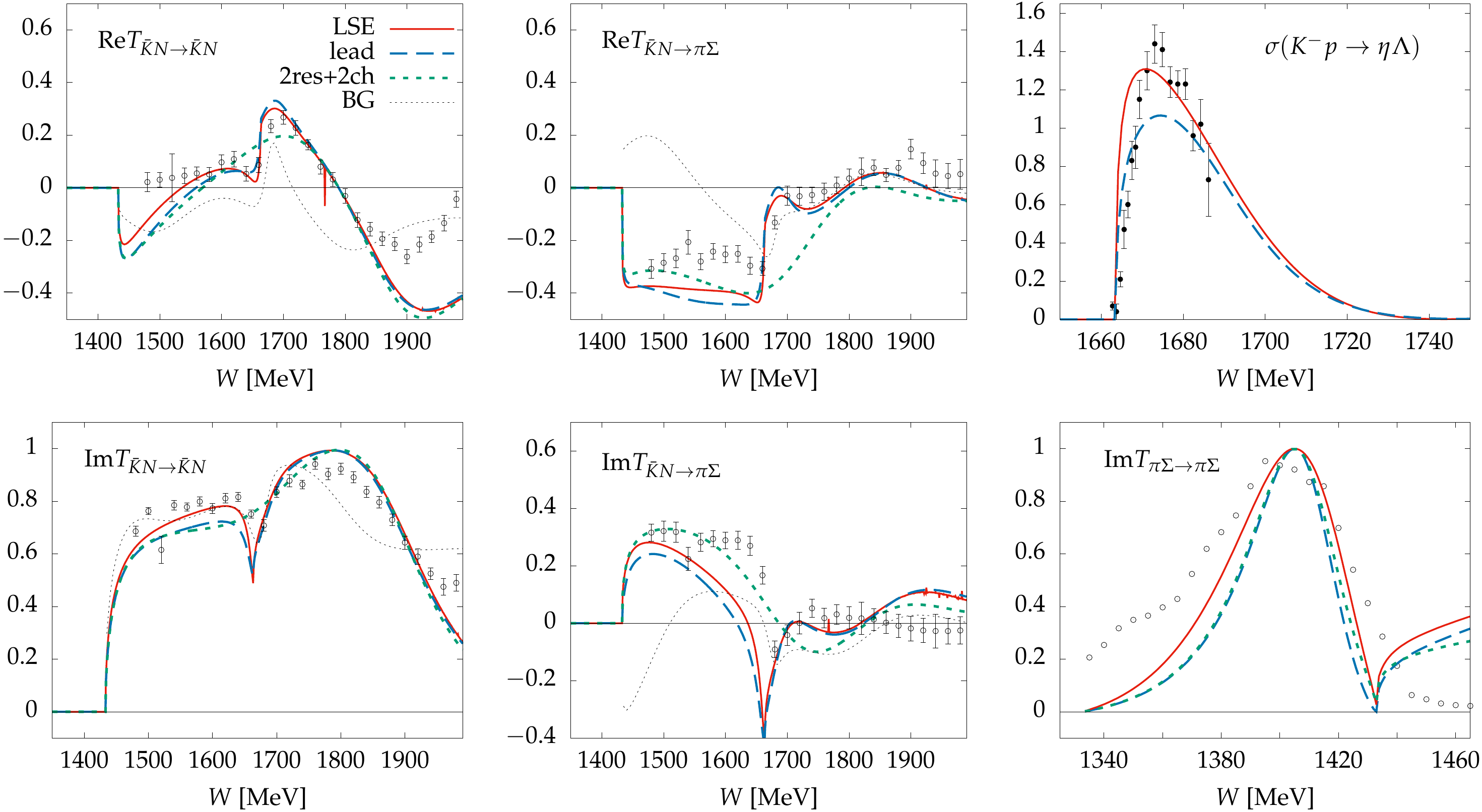}
\end{center}
\vspace{-12pt}
\caption{The scattering amplitudes for reactions $\bar{K}N\to\bar{K}N$,
$\bar{K}N\to\pi\Sigma$ and $\pi\Sigma\to\pi\Sigma$, and the 
cross-section for $K^-p\to\eta\Lambda$.
``LSE'' stands for solving the Lippmann-Schwinger equation, 
``lead'' for the leading order, 
``2res+2ch'' for only two channels and two resonances, 
``BG'' is the Bonn-Gatchina solution~\cite{Sarantsev19a}.
Experimental data from~\cite{Manley13a} for the amplitudes, 
\cite{BNL} for the cross-section 
and \cite{Hemingway1985} for the invariant mass distribution.}
\label{fig:multi-0}
\end{figure*}

As we have mentioned in the previous section, using the strength
of the contact interaction that produces two poles below the
$KN$ threshold results in the scattering amplitudes which strongly 
disagree with experiment.
Only when the contact interaction is reduced to less than 
10~\% of that strength, the calculated amplitudes start to 
exhibit the typical pattern seen in the $\bar{K}N\to\bar{K}N$ 
and $\bar{K}N\to\pi\Sigma$ reactions.
In fact, we have found the best agreement by putting the
strength of the contact interaction to zero.
This finding agrees with the results of the analysis of Kamano 
et al.~(\cite{Kamano14}, Model A) who find a good overall agreement 
in this partial wave without including the contact interaction.

The results for scattering in the $S_{01}$ partial wave are displayed 
in fig.~\ref{fig:multi-0}; the real and the imaginary parts of 
the $T$ matrix are compared to the results for $\bar{K}N\to\bar{K}N$ 
and  $\bar{K}N\to\pi\Sigma$ from the single-energy partial-wave 
analysis \cite{Manley13a}, as well as to the analysis of 
the Bonn-Gatchina group~\cite{Sarantsev19a}.
Furthermore, the calculated cross-section for $K^-p\to\eta\Lambda$ 
is compared to the measured one~\cite{BNL}, and in addition,
the imaginary part of $T(\pi\Sigma\to\pi\Sigma)$ is confronted
with the $\pi\Sigma$ invariant mass spectrum~\cite{Hemingway1985}.
The bare mases used in the calculation are displayed in 
table~\ref{tab:poles}; we assume no mixing between the two bare 
octet configurations, the $K\Xi$ couplings to all $\Lambda^*$ are 
reduced to 30~\% of the QM value, while the $\eta_8\Lambda$ 
coupling to the octet $\Lambda^*$ even to 10~\% of the QM value.

\begin{table}[h]
\caption{\label{tab:poles}
The pole parameters obtained from the $S_{01}$ partial wave amplitudes 
in different channels using the Laurent-Pietarinen expansion;
``lead'' stands for the solution in the leading order; 
$m^0$ is the bare quark mass.}
{\renewcommand{\arraystretch}{1.5}
\setlength{\tabcolsep}{3pt}       
\begin{tabular}{lcrrrcc} 
\hline
res. & chan. & ~~Re$W$ & $-2{\rm Im}W$ & modul. &  QM & $m^0$ \\
    &       & [MeV] & [MeV]         & [MeV] &  assign. & [MeV]  \\
\hline
$\Lambda(1405)$& $\pi\Sigma$-$\pi\Sigma$& 1417 & 30 & 14.4 & $^21[70]$ & 1667\\
              &  (lead)                 & 1413 & 26 & 13.1 &           & 1637\\
\hline
$\Lambda(1670)$& $\pi\Sigma$-$\pi\Sigma$& 1667 & 29 & 7.3 & $^48[70]$  & 1720\\
               &    (lead)              & 1669 & 32 & 9.5 &            & 1713\\
               & $\bar{K}N$-$\bar{K}N$  & 1664 & 26 & 3.8 &  &     \\
               &     (lead)             & 1664 & 34 & 4.8 &  &     \\
               & $\bar{K}N$-$\pi\Sigma$ & 1664 & 36 & 6.7 &  &     \\
               &     (lead)             & 1666 & 43 & 9.0 &  &     \\
\hline
$\Lambda(1800)$& $\bar{K}N$-$\bar{K}N$  & 1885 & 341 & 157 & $^28[70]$ & 1746\\ 
               &     (lead)             & 1882 & 358 & 204 &           & 1749\\ 
               & $\bar{K}N$-$\pi\Sigma$ & 1814 & 198 &  26 &  &      \\
               &     (lead)             & 1790 & 207 &  31 &  &      \\ 
\hline
\end{tabular}}  
\end{table}

Our calculation shows that the scattering amplitudes are
dominated by the resonant terms and that the
background potential plays a rather minor role.
Furthermore, evaluating the amplitudes without solving the LSE,
i.e. by keeping only the leading terms in Eqs.~(\ref{eq4VR}) 
and (\ref{eq4D}), and readjusting sightly the bare masses,
the result of the full calculation changes only insignificantly 
(long dashes vs. solid line for the LSE in fig.~\ref{fig:multi-0}).

We have performed the Laurent-Pietarinen expansion to determine the 
positions of the poles in the complex $W$ plane (table~\ref{tab:poles}).
The mass and the width of the lowest pole determined in the $\pi\Sigma$
channel are consistent with the PDG values for the $\Lambda(1405)$.
The values for the second pole are calculated from the amplitudes
for three different reactions, and all three give values consistent
with the PDG result for the $\Lambda(1670)$.
For the third pole the values from $\bar{K}N$-$\bar{K}N$ differ
more substantially from the preferred PDG values, while the results
for $\bar{K}N$-$\pi\Sigma$ agree well with the PDG values.
These results change only marginally in the leading order.

In our approach only the $\pi\Sigma$ channel provides 
the information about the poles below the $KN$ threshold. 
Still, some information can be obtained 
also from the $\bar{K}N$ amplitude by expanding
the $T$ matrix for small $k$~\cite{Kamano15},
$
  T = k\left(a^{-1} -i k + {1\over2} r_ek^2\right)^{-1}
$,
which yields the scattering length
$a_{0\,\mathrm{LSE}}=(-1.32 + i\,1.00)$~fm
and $a_{0\,\mathrm{lead}}=(-1.42 + i\,0.71)$~fm,
and the pole at $W_{\mathrm{LSE}} = (1434.8 - i\,22.1)$~MeV
and at $W_{\mathrm{lead}} = (1425.6 - i\,21.5)$~MeV.
Both values of the scattering length are inside the 
allowed region established in Ref.~\cite{Doering11} and deduced 
from the SIDDHARTA measurements~\cite{SIDDHARTA}.
While the width is consistent with the values in table~\ref{tab:poles}
for the $\pi\Sigma$ channel, the mass comes much too close
to the threshold value  in this approximation.

Next we have considered the case with only two channels, 
$\pi\Sigma$ and $\bar{K}N$, and two bare three-quark states, 
the $\Lambda_1^*$ and ${}^2\Lambda_8^*$.
Again we obtain a similar behaviour of amplitudes as in the 
full calculation in a wide energy range except for the 
interval in the vicinity of the second resonance $\Lambda(1670)$,
neglected in this approximation.

We shall discuss the nature of the lowest resonance
in Sec.~\ref{sec:structure}.

Let us comment here on similar calculations in the framework of the 
same model in  Refs.~\cite{Thomas85} and \cite{Fink90} where the 
contact interaction as well as the bare three-quark state were included.
In the former work a single pole below the $KN$ threshold was found, 
while in the latter two poles were found in the  $\pi\Sigma$ channel, 
one close to the $KN$ and another close to the $\pi\Sigma$ threshold.
In both approaches the choice of the bag radius and pion decay constant
resulted in a rather weak strength of the contact interaction as well as
of the quark-meson interaction, hence 
the resonances below the $KN$ threshold were generated through
the dynamical mechanism discussed in the previous subsection.

\subsection{Including three-quark states in the $S_{11}$ partial wave}

In the $S_{11}$ partial wave we include five channels, $\pi\Sigma$,
$\bar{K}N$, $\pi\Lambda$, $K\Xi$ and $\eta\Sigma$, and two
bare quark state corresponding to $^2\Sigma^*_8$ and $^4\Sigma^*_8$.
The inclusion of the contact interaction turns out to be mandatory here.
With our standard choice for $R$ and $f_\pi$ 
controlling the meson-baryon interaction, 
a satisfactory agreement --- at least for $W$ below 1750~MeV --- 
is reached by using $f_\pi=128$~MeV for the contact interaction.
In addition, we assume that the strength of the $\pi\Sigma\Sigma^*_8$
coupling constant is reduced by 30~\% with respect
to the quark-model value,  and a mixing angle of $20^\circ$
is used already at the level of bare $^2\Sigma^*_8$ and $^4\Sigma^*_8$ 
states.
For the channels $\bar{K}N$, $\pi\Lambda$, $K\Xi$ we fix the coupling 
constants to their quark-model values, while for the  $\eta\Sigma$
channel we use the same prescription as for the $\eta\Lambda$ channel
in the $S_{01}$ partial wave.
The optimal masses of the bare quark states remain close to their 
nominal values 1750~MeV and 1900~MeV, i.e. 
$m(^2\Sigma^*_8)=1750$~MeV and $m(^4\Sigma^*_8)=1876$~MeV.

\begin{figure*}[h]
\begin{center}
\includegraphics[width=170mm]{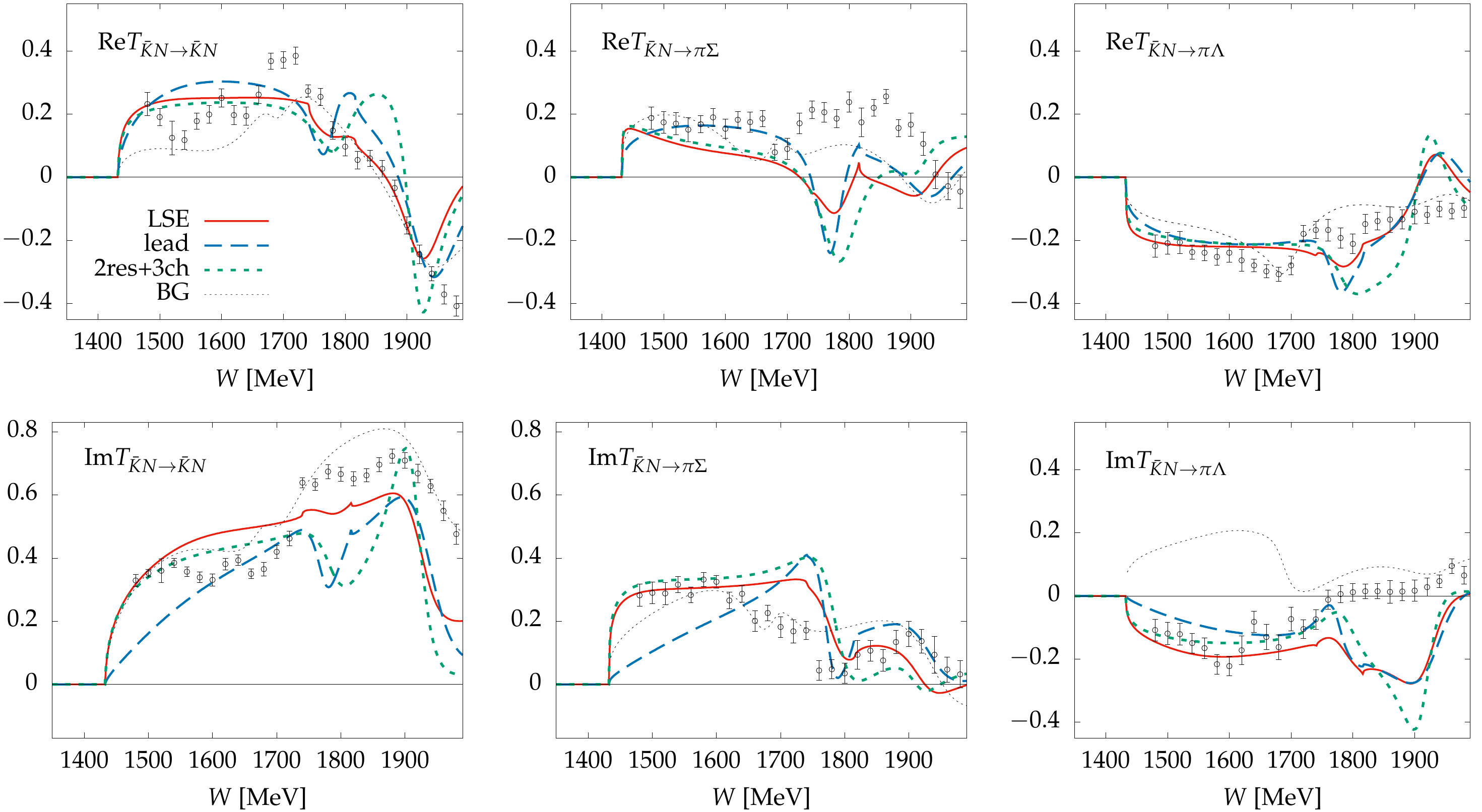}
\end{center}
\vspace{-12pt}
\caption{Amplitudes for $I=1$: the solid line represents 
the full solution, the dashes the leading order solution,
and the short dashes the full solution with only three channels.
The tiny dashes and the experimental points as in fig.~\ref{fig:multi-0}.}
\label{fig:multi-1}
\end{figure*}

In fig.~\ref{fig:multi-1} we compare the full calculation  by
solving LSE with all five channels, the full calculation which 
includes only $\pi\Sigma$, $\bar{K}N$, $\pi\Lambda$ channels, and 
the calculation in the leading order (without solving LSE).
As expected, the first three channels dominate at lower $W$,
but in contrast to the $S_{01}$ partial wave, the  leading order 
solution differs considerably from the full solution as a 
consequence of a much stronger potential that enters the LSE.

The positions of the poles in the complex $W$ plane are displayed in
table~\ref{tab:poles1}. 
While the lower pole is located at too large Re$W$ and too small Im$W$ 
compared to the PDG values, the upper pole is better reproduced
in the case of five channels.
The scattering length in the case of five (three) channels is
$a_1=(0.54 -i\, 0.39)$~fm  ($a_1=(0.52 - i\, 0.41)$~fm); 
while the real parts are well within the allowed region
advocated in Ref.~\cite{Doering11}, the imaginary parts seems 
to be slightly too low.

\begin{table}[h]
\caption{\label{tab:poles1} 
The pole parameters obtained in the $S_{11}$ partial wave by using five or 
three channels.}
\bigskip
{\renewcommand{\arraystretch}{1.5} 
\setlength{\tabcolsep}{3pt}        
\begin{tabular}{lcrrrcc} 
\hline
res. & chan. & ~~Re$W$ & $-2{\rm Im}W$ & modul. &  QM & $m^0$ \\
react. &       & [MeV] & [MeV]         & [MeV] &  assign. & [MeV]  \\
\hline
 $\Sigma(1750)$           &5 \& 3&      &     &      & $^28[70]$  &1750\\
 $\bar{K}N$-$\bar{K}N$    & 5    & 1738 &  48 &  1.7 &  &     \\
                          & 3    & 1784 &  75 &  7.8 &  &     \\
 $\bar{K}N$-$\pi\Sigma$   & 5    & 1786 &  54 &  4.8 &  &     \\
                          & 3    & 1785 &  75 & 13.8 &  &     \\
 $\bar{K}N$-$\pi\Lambda$  & 5    & 1788 &  49 &  2.1 &  &     \\
                          & 3    & 1785 &  73 &  5.1 &  &     \\ 
\hline
$\Sigma(1900)$            &5 \& 3&      &     &      & $^48[70]$ &1876\\
$\bar{K}N$-$\bar{K}N$     & 5    & 1924 &  96 & 18.1 &  &      \\ 
                          & 3    & 1914 &  61 & 21.5 &  &      \\ 
$\bar{K}N$-$\pi\Sigma$    & 5    & 1925 & 123 & 10.7 &  &      \\
                          & 3    & 1914 &  60 &  3.0 &  &      \\ 
$\bar{K}N$-$\pi\Lambda$   & 5    & 1924 &  81 & 10.3 &  &      \\
                          & 3    & 1914 &  61 & 12.9 &  &      \\ 
\hline
\end{tabular}}  
\end{table}

\begin{figure}[h]
\begin{center}
\includegraphics[width=70mm]{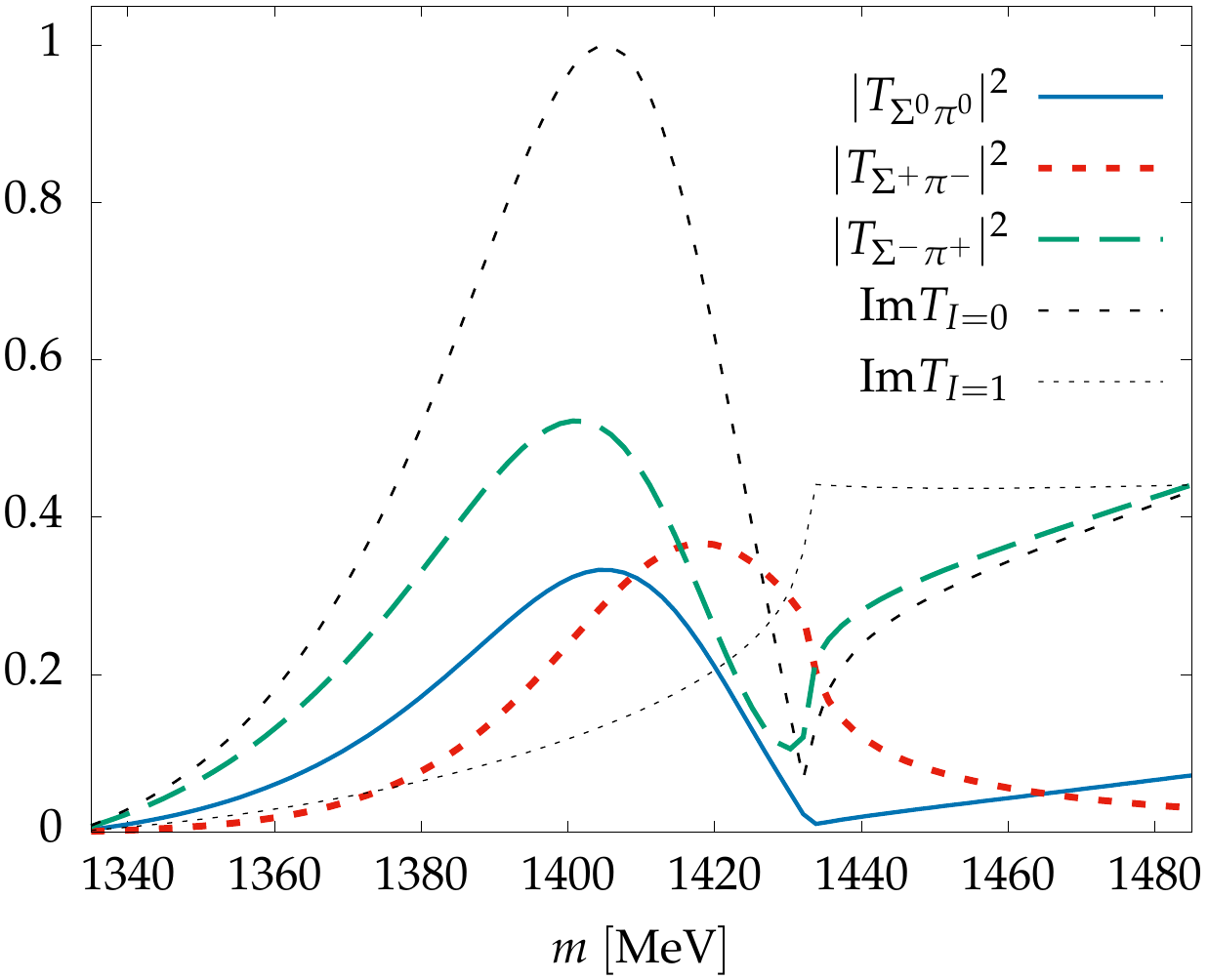}
\end{center}
\vspace{-12pt}
\caption{The (squared) $T$ amplitudes for $\Lambda^*\to\Sigma\pi$ 
as function of the invariant mass $m$.} 
\label{fig:TSpi2}
\end{figure}

From both partial waves we can construct the amplitudes for the decay 
of the $\Lambda(1405)$ into $\Sigma^+\pi^-$, $\Sigma^-\pi^+$, and 
$\Sigma^0\pi^0$, and compare them to those extracted in 
the reactions $p + p \to \Sigma^{\pm} + \pi^{\mp} + K^+ + p$
by the HADES collaboration~\cite{HADES13}, and 
$\gamma + p \to K^+ + \Sigma + \pi$ by the CLAS Collaboration~\cite{CLAS13}.
The corresponding $|T_{\Sigma^-\pi^+}|^2$,  $|T_{\Sigma^0\pi^0}|^2$ and
$|T_{\Sigma^+\pi^-}|^2$ can be straightforwardly expressed in terms 
of the $T$-matrix amplitudes for the $S_{01}$ and $S_{11}$ partial waves
(assuming no contribution from $I=2$) and related to the 
cross-section for the above reactions.
We shall not attempt to write down the explicit expression for
the cross-section but rather compare the qualitative behaviour of
$|T_{\Sigma\pi}|^2$ with the mass distribution of the model by 
Bayar et al.~\cite{Bayar18} based on HADES data.
Comparing our fig.~\ref{fig:TSpi2} with their fig.~6 we notice that the 
positions of peaks for the $\pi^+$,  $\pi^0$ and  $\pi^-$ distributions 
are similar, also the $\pi^+$ distribution is dominant in both cases.
Let us note that below the $KN$ threshold the 
$\pi^0$ distribution is one third of the Im$T_{\pi\Sigma-\pi\Sigma}$
amplitude and is in our case peaked around the nominal
mass of the resonance at $m=1405$~MeV which corresponds 
to one of the free parameters in our model.  Such a value 
is supported also by the analysis of HADES data 
by Hassanvand et al.~\cite{Hassanvand13}.  A similar
comparison of mass distributions in different channels
to those obtained by \cite{Ulf15} and \cite{Mari21} 
(both based on CLAS data), is less conclusive.

\section{\label{sec:structure} The structure of the resonances}

\subsection{Evolution of poles in the complex energy plane}

In order to obtain a deeper insight into the mechanism of resonance 
formation in the presence of genuine three-quark states
in the $S_{01}$ partial wave we follow the evolution of the 
$S$-matrix poles in the complex energy plane as a function of the 
interaction strength by performing the Laurent-Pietarinen expansion.
We start from the genuine three-quark states and calculate the 
scattering amplitudes by gradually increasing the  meson-baryon 
coupling constants in all channels by a factor $g,\> 0< g\le 1$, 
to finally reach the physical values used in the previous Section.
This approach has been used in our previous work~\cite{PRC2018,PRC2019} 
to show that the Roper resonance evolves from the genuine three-quark 
state, while the $\Delta(1600)$ emerges as a dynamically generated state.

\begin{figure*}[t]
\begin{center}
\includegraphics[width=170mm]{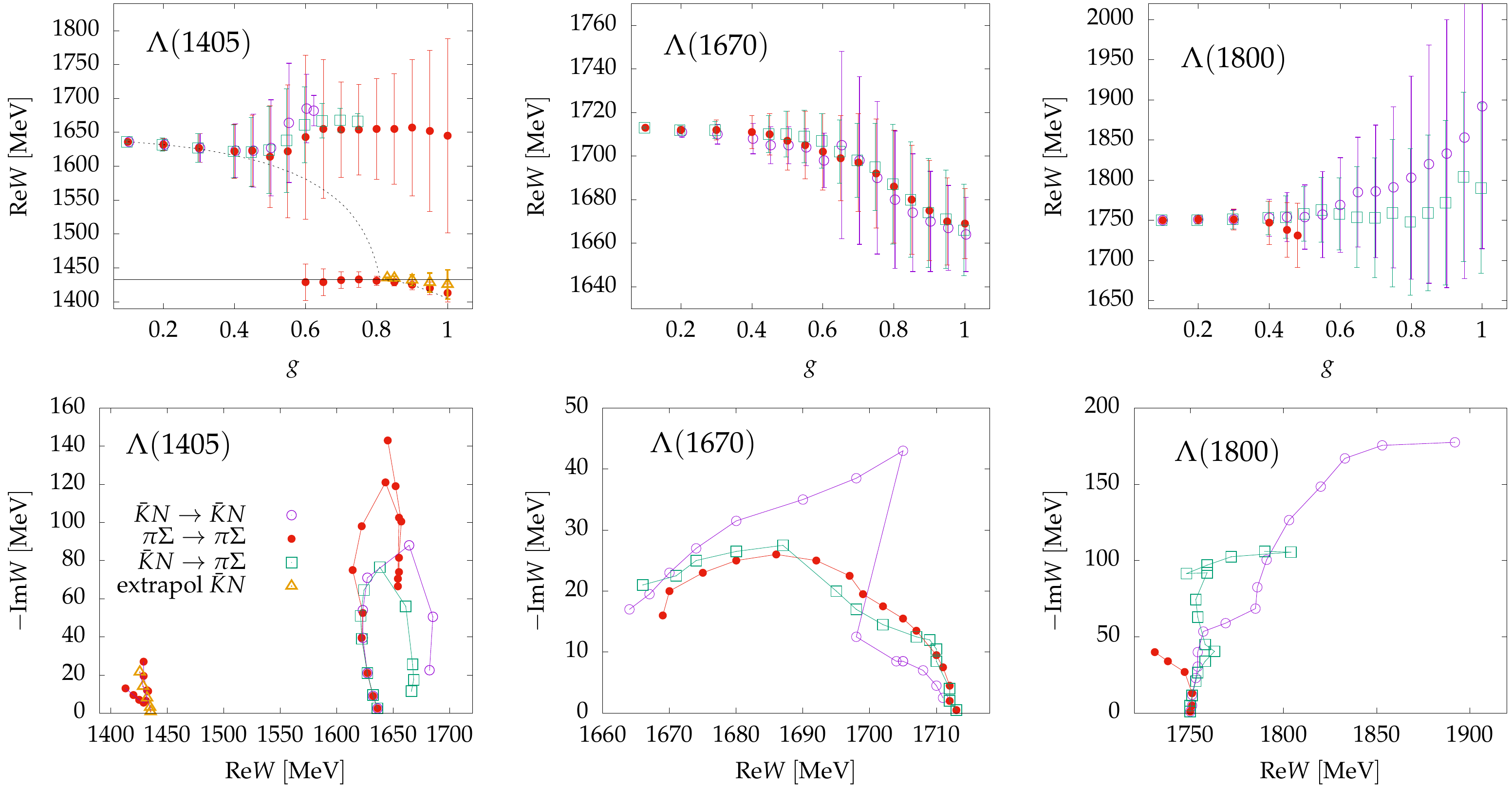}
\end{center}
\vspace{-12pt}
\caption{Evolution of poles of the three resonant states as a function 
of interaction strength (top panels) with error bars corresponding 
to $\pm|\mathrm{Im}W|$, and in the complex $W$ plane (bottom panels). 
The full circles correspond to $\pi\Sigma\to\pi\Sigma$, the empty circles 
to $\bar{K}N\to\bar{K}N$, squares to $\bar{K}N\to\pi\Sigma$,
and triangles to the $\bar{K}N\to\bar{K}N$ amplitudes extrapolated
below the threshold.}
\label{fig:poles}
\end{figure*}

In the present case we deal with an evolution of three resonances 
that lie relatively close to each other and strongly mix, 
particularly in the region of intermediate coupling strengths.
This represent a serious difficulty in identifying the poles
belonging to individual resonances, which may overlap or even cross.
Furthermore, the presence of the $KN$ and $\eta\Lambda$
thresholds may strongly influence the poles in their vicinity.
It turns out that the procedure is more stable if one uses the 
leading solution due to smaller widths of the resonances.
Yet the final solution stays close to the full calculation
(see table~\ref{tab:poles}), so the conclusion should be valid
also for the full solution.

The evolution for the reactions $\bar{K}N\to\bar{K}N$,
$\bar{K}N\to\pi\Sigma$ and $\pi\Sigma\to\pi\Sigma$ is
shown in fig.~\ref{fig:poles}.
We do not show $\bar{K}N\to\eta\Lambda$ since the relevant
pole lies very close to the threshold and its determination is less
reliable.

The evolution of the lower resonance (left panels) starts with 
the three-quark singlet configuration.
The evolution for $\bar{K}N\to\bar{K}N$ and $\bar{K}N\to\pi\Sigma$ 
stops at $g=0.6$ and $g=0.7$, respectively, as the widths 
and moduli vanish.
This does not happen in the $\pi\Sigma$ channel, 
the evolution rather continues away from the real axis.
Beyond $g>0.5$ another branch appears and evolves toward 
the pole that can be identified as $\Lambda(1405)$; 
at larger $g$ this branch can also be obtained 
by using the small-$k$ expansion for $T_{\bar{K}N\,\bar{K}N}$.
From our analysis it is unclear whether 
(i) this branch  emerges at the threshold and evolves independently
of the upper branch or
(ii) it smoothly evolves from the genuine three-quark state, in which 
case there would exist a bifurcation for $\pi\Sigma\to\pi\Sigma$
in the intermediate regime of $g$.
Though the curves presented in fig.~\ref{fig:poles} favor
the first possibility, we should mention that the determination
of the pole in the vicinity of the threshold is unreliable
and its position very close to the threshold may be an artifact.
Plotting the (real) pole of the $K$ matrix as a function of $g$
supports the second possibility since it exhibits a rather rapid
transition from the bare value to values below the threshold.

The branch above $g=0.80$, where the mass of the pole starts
moving away from the threshold value, has a clear physical 
interpretation, which will be discussed in the following.   

The evolution of the middle resonance (central panels) starting 
with the three-quark octet configuration with spin $S=\thalf$ 
is smooth except around $g=0.6$ in the $\bar{K}N$  channel where 
the pole pertaining to the $\Lambda(1405)$ (left panel) 
in this channel disappears.
All three evolutions end up at the pole which can unmistakably
be attributed to the physical $\Lambda(1670)$ resonance,
and confirm the assignment given in table~\ref{tab:poles}. 

The evolution of the upper resonance (right panels) starting with 
the three-quark octet configuration with spin $S=\half$ is smooth
and in $\bar{K}N\to\bar{K}N$ and $\bar{K}N\to\pi\Sigma$ evolves to 
the resonance that can be identified as $\Lambda(1800)$, though in 
the $\bar{K}N$ channel it terminates at too large Re$W$ and Im$W$.
The $\pi\Sigma$ system is only weakly coupled to the bare state 
and above $g=0.5$ becomes too weak to be detected.

\subsection{Structure of the $\Lambda(1405)$ resonance}

\begin{figure}[h]
\begin{center}
\includegraphics[width=70mm]{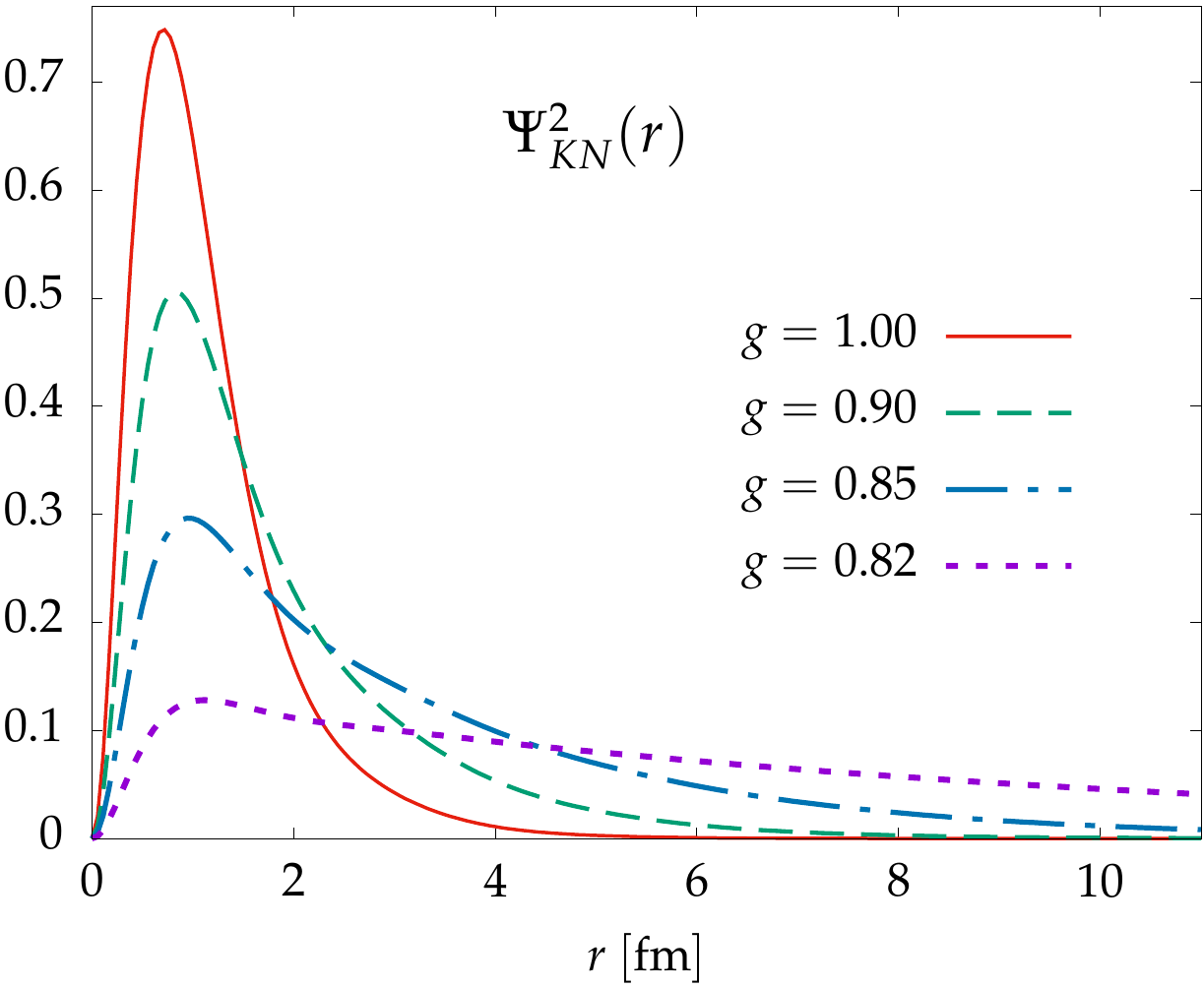}
\end{center}
\vspace{-12pt}
\caption{Kaon probability density for different 
relative coupling strengths.} 
\label{fig:Kwf}
\end{figure}

\begin{figure}[h]
\begin{center}
\includegraphics[width=80mm]{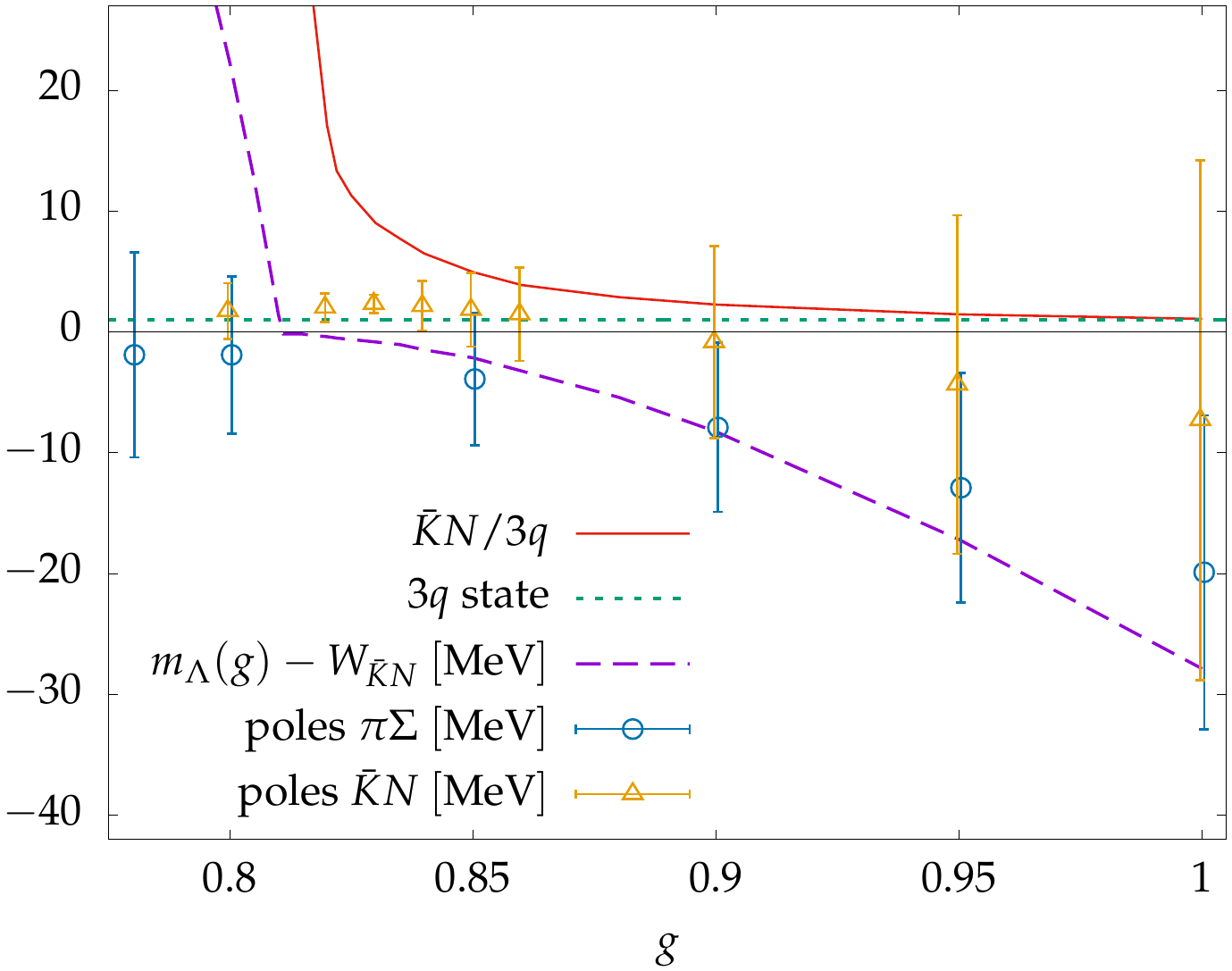}
\end{center}
\vspace{-12pt}
\caption{Evolution of the $\Lambda(1405)$ (Breit-Wigner) mass 
(dashes), pole mass and width (error bars) 
from the $\pi\Sigma$ (circles) and $\bar{K}N$ (triangles) channels,
and the weight of the $\bar{K}N$ component (solid line) normalized such 
that the weight of the three-quark state is equal one (short dashes).
Masses are measured with respect to the $KN$ threshold.}
\label{fig:KNweight}
\end{figure}

Let us observe the evolution of the $\bar{K}N$ channel 
below the $KN$ threshold as $W$
approaches $m_\Lambda$, the lowest pole of the $K$ matrix.
If we normalize the corresponding channel state (\ref{PsiH})
by inverting the norm (\ref{normPV}) and taking into account 
that both terms are dominated by 
$c_{\bar{K}N\Lambda^*}\propto (W-m_{\Lambda})^{-1}$,
the channel state can be cast in the form
\begin{eqnarray}
|\Psi_{\bar{K}N}\rangle &=& \sqrt{Z}\left(|\Phi^0_{\Lambda^*}\rangle
  + \int{dk\,\mathcal{V}_{\bar{K}N\Lambda^*}(k)\over \omega_k+E_N(k)-W}
     \left[a_K^\dagger(k)|\Phi_N\rangle\right]\right.
\nonumber\\
  &+& \left.\int{dk^\pi\,\mathcal{V}_{\pi\Sigma\Lambda^*}(k^\pi)
       \over \omega_k^\pi+E_\Sigma(k^\pi)-W} 
       \left[a_\pi^\dagger(k^\pi)|\Phi_\Sigma\rangle\right]
  + \ldots\right)\,,
\label{psiKN}
\end{eqnarray}
where $k$ refers to the kaon momentum.
The terms involving the $\eta\Lambda$ and $K\Xi$ components, 
as well as the octet admixtures to singlet three-quark state, 
are small and will be neglected in the following.
The norm then reads
\begin{equation}
Z^{-1} = 1 + \int {dk\,\mathcal{V}^2_{\bar{K}N\Lambda^*}(k)
     \over (\omega_k+E_N(k)-W)^2}
  - {d\over dW} \Sigma_{\pi\Sigma}(W)\,.
\label{Znorm}
\end{equation}
We notice that for $W$ close to the threshold, the second term in 
Eq.~(\ref{Znorm}) strongly dominates and the system is very loosely
bound (see fig.~\ref{fig:Kwf}), however, at the physical 
value of the coupling strength it becomes comparable to the weight 
of the bare three-quark component  (see fig.~\ref{fig:KNweight}).
The contribution from the $\pi\Sigma$ component (expressed in terms 
of the derivative of the self-energy) remains very small.
On the other hand, the contribution to the energy is
dominated by both the $\bar{K}N$ as well as the $\pi\Sigma$ 
self-energies, which are responsible to push the mass of the
physical $\Lambda(1405)$ below the $KN$ threshold; this mechanism 
is similar to the one proposed by Arima et al.~\cite{Arima94}.

Our model therefore confirms the picture in which the $\Lambda(1405)$ 
is predominantly a $\bar{K}N$ molecular state; however,
in our model the binding mechanism is not the contact interaction 
that would generate attraction between the (anti)kaon and the nucleon
but rather the $\bar{K}N\Lambda^*$ interaction which implies 
the presence of a bare three-quark configuration with the quantum 
numbers of the resonance; the presence of the $\pi\Sigma$ channel
is also necessary to ensure the binding.
Let us mention that the $\Lambda(1405)$ is not a Feshbach resonance
since the energy of the state~(\ref{psiKN}) alone is well above
the $KN$ threshold.

A similar model with a bare state and the kaon cloud around
the nucleon was proposed long ago by Thomas et al.~\cite{Thomas85}
in the framework of the same model;
in our approach we have extended the model by inclusion of other
channels and resonances, but also showing that the presence of the
contact term is not mandatory.
Our picture of the resonance can also be related to the state 
found on the lattice~\cite{Leinweber12,Leinweber15} and interpreted 
in the framework of Hamiltonian effective theory \cite{Leinweber18}.

\section{Conclusion}

As we have shown in Sec.~\ref{sec:scattering} our model is able 
to generate a $\half^-$ resonance --- or even two resonances ---
below the $KN$ threshold either as a molecular state 
or a genuine three-quark state dressed with baryon-meson pairs.
We believe that in order to give credence to either of
the two approaches
it is important to carry out the calculation of the $S_{01}$ partial 
wave amplitudes in the relevant channels by treating all three 
resonances in a unified framework.
The main and, admittedly, rather surprising, conclusion of our 
investigation is that the scattering amplitudes 
are dominated by the quark degrees of freedom rather than by 
non-linear dynamics of the baryon-meson systems.
By using our standard choice of model parameters 
which had successfully reproduced the scattering as well as 
electroproduction amplitudes in the nonstrange sector we have been 
able to obtain a satisfactory result already in the leading order; 
solving the LSE only marginally improves the results.
Furthermore, as in the calculation of Kamano et al. \cite{Kamano14},
we have found that including the contact 
interaction does not improve the results.
Our results therefore confirm that the main mechanism to lower 
the mass of the $\Lambda(1405)$ by $\approx 250$~MeV with respect to 
its bare value is the one suggested by Arima et al. \cite{Arima94},
that is, the attractive self-energy term in the $\pi\Sigma$ and 
$\bar{K}N$ channels, with the latter term being strongly enhanced 
due to the presence of the  $KN$ threshold.
Nonetheless, even without including the contact interaction, we have 
been able to observe the formation of the $\bar{K}N$ molecular state.
By gradually increasing the strength of all meson-baryon couplings, 
a resonant state, strongly dominated by a weakly bound $\bar{K}N$ 
component and an almost negligible admixture of the bare three-quark 
and $\pi\Sigma$ components, emerges slightly below the $KN$ threshold.
At the physical strength, at which this state can be identified with 
the $\Lambda(1405)$ resonance, the molecular component becomes weaker
but still dominates over the bare quark state which, in turn, 
is dominated by the singlet component.
There is, however, an important difference between our state
and the molecular state  of the chiral unitary approach;
in our approach the attraction is generated through the
$\bar{K}N\Lambda^*$ coupling and therefore the presence of the
singlet state is mandatory.

Regarding the two higher lying resonances, the $\Lambda(1670)$ 
is well reproduced and assigned to $S={3\over2}$;
with the $\Lambda(1800)$ there remains some ambiguity about 
the determination of its mass in different channels, which 
signals that our model becomes less reliable at higher $W$.

In the $S_{11}$ partial wave it turns out that the inclusion of 
the contact interaction is important; we reproduce
reasonably well the scattering amplitudes in the energy range
from the threshold up to around 1750 MeV using either all five 
channels or solely the $\pi\Sigma$, $\bar{K}N$, and $\pi\Lambda$
channels.

\appendix

\section{\label{vertices}The Cloudy Bag Model meson-quark vertices
and coupling constants}

The $s$-wave quark-meson vertices $\hat{V}(k)$ in Eq.~(\ref{qM}) are 
evaluated in the Cloudy Bag Model assuming that in the resonant 
state one of the three quarks is excited from the $1s$ state to 
the $1p_{1/2}$ state. 

For the quark part of the quark-pion, quark-eta meson and quark-kaon 
interaction we obtain
\begin{eqnarray}
\hat{V}^\pi_{l=0,t}(k) &=& 
     V^\pi(k)\sum_{i=1}^3 \tau_t(i)\, \mathcal{P}_{sp}(i)\,,
\label{Vpi}\\
   \hat{V}^\eta(k) &=&V^\eta(k) \sum_{i=1}^3
          \lambda_8(i)\,\mathcal{P}_{sp}(i)\,,
\label{Veta}\\
   \hat{V}^K_{t}(k) &=& V^K(k)
\label{VK}
\end{eqnarray}
Here $\mathcal{P}_{sp} =\sum_{m_j}|sm_j\rangle\langle p_{1/2}m_j|$,
$\omega_s=2.043$, $\omega_{p_{1/2}}=3.811$,
$X(K^0)= -(\lambda_6- i\lambda_7)/\sqrt2$,
$X(\bar{K}^0)= -(\lambda_6+i\lambda_7)/\sqrt2$,
$X(K^+)= -(\lambda_4- i\lambda_5)/\sqrt2$,
$X(K^-)=  (\lambda_4+ i\lambda_5)/\sqrt2$.
Assuming $f_\eta=f_K=f_\pi$, the form-factors of the surface part
and of the volume part take the form
\begin{eqnarray}
    V^\pi_S(k) &=& V^\eta_S(k) =  V^K_S(k)
\nonumber\\
              &=& {1\over2f_\pi}\sqrt{\omega_{p_{1/2}}\omega_s\over 
    (\omega_{p_{1/2}}+1)(\omega_s-1)}\,
     {1\over2\pi}\,{k^2\over\sqrt{\omega_k}}\,{j_0(kR)\over kR}\,,
\nonumber\\
V_V(k) &=& \left[{\omega_s\over R} - {\omega_{p1/2}\over R} + \omega_M(k)\right]
\nonumber\\ &\times&
  \int_0^R dr \,r^2[u_s(r)u_p(r)+v_s(r)v_p(r)]j_0(kr)\,.
\end{eqnarray}

For the physical $\eta$ we assume 
$\eta =  \cos\theta_P\eta_8-\sin\theta_P\eta_1 $, 
for the singlet $\eta_1$ $\lambda_8$ is replaced by $\lambda_1$
in Eq.~(\ref{Veta}).

The coupling constants for the $s$-channel exchange potential are 
collected in table~\ref{tab:gs}, those for the $u$-channel exchange 
potential involving strange baryons in table~\ref{tab:gus}, and 
for exchange of nonstrange baryons  in table~\ref{tab:gun}.

\smallskip
\noindent
\begin{table}[h]
\caption{\label{tab:gs}
Reduced $X^*\to MB$ matrix elements}
{\renewcommand{\arraystretch}{2.0}
\setlength{\tabcolsep}{3pt}       
\begin{tabular}{lrrrrrrrr} 
\hline
$X^*/BM$
      & $\pi\Sigma$ 
      & $\pi\Lambda$ 
      & $\eta_1\Lambda$ 
      & $\eta_8\Lambda$ 
      & $\eta_1\Sigma$ 
      & $\eta_8\Sigma$ 
      & $\bar{K}N$
      & $K\Xi$\cr
\hline
$\Lambda^*_1$           & $\sqrt3$          & 0 & 0  & $-1$ & 0 & 0
                        & $\sqrt2$ & $\sqrt2$ \cr
$\Lambda^*_{28}$  & $ -{\sqrt3\over3}$ & 0 & ${2\sqrt2\over3}$ &$-{1\over3}$
                       & 0 & 0
                       & $\sqrt2$ & $-{2\sqrt2\over3}$ \cr
$\Lambda^*_{48}$ & ${2\sqrt3\over3}$ & 0 & ${2\sqrt2\over3}$ &${2\over3}$
                       & 0 & 0
                       & $0$  & $-{2\sqrt2\over3}$ \cr
$\Sigma^*_{28}$   & $-{5\over3}\sqrt{2\over3}$  
                       &  $-{1\over3}$ 
                       & 0 & 0 
                       & ${2\sqrt{2}\over3}$
                       & ${1\over3}$
                       & $-{1\over3}\sqrt{2\over3}$
                       & ${4\over3}\sqrt{2\over3}$ \cr
$\Sigma^*_{48}$  & $-{2\over3}\sqrt{2\over3}$  
                       & ${2\over3}$ 
                       & 0 & 0 
                       & ${2\sqrt{2}\over3}$ 
                       & $-{2\over3}$ 
                       & $-{4\over3}\sqrt{2\over3}$ 
                       & $-{2\over3}\sqrt{2\over3}$ \cr
\hline
\end{tabular}}  
\end{table}
\smallskip

\noindent
\begin{table}[h]
\caption{\label{tab:gus}
Reduced $B\to MX^*$ matrix elements}
{\renewcommand{\arraystretch}{2.0} 
\setlength{\tabcolsep}{3pt}        
\begin{tabular}{lrrrrrrrr} 
\hline
$X^*/BM$ & $\Sigma\pi$ 
      & $\Lambda\pi$ 
      & $\Lambda\eta_1$
      & $\Lambda\eta_8$
      & $\Sigma\eta_1$ 
      & $\Sigma\eta_8$ 
      & $ N    K$
      & $\Xi   \bar{K}$\cr
\hline
$\Lambda^*_1$           & $1$ & 0  & 0  & $-1$ & 0 & 0 & 1 & 1 \cr
$\Lambda^*_{28}$  & ${1\over3}$ & 0 & ${2\sqrt2\over3}$ & $-{1\over3}$ 
                        & 0 & 0 
                        & 1  & $-{2\over3}$ \cr
$\Lambda^*_{48}$ & $-{2\over3}$  & 0 & ${2\sqrt2\over3}$ & ${2\over3}$
                        & 0 & 0
                        & 0 & $-{2\over3}$ \cr
$\Sigma^*_{28}$   & $-{5\over3}\sqrt{2\over3}$  
                       & ${\sqrt3\over3}$
                       & 0 & 0 
                       & ${2\sqrt2\over3}$ & ${1\over3}$
                       &     $-{1\over3}$  
                       & ${4\over3}$ \cr
$\Sigma^*_{48}$  & $-{2\over3}\sqrt{2\over3}$  
                       & $-{2\sqrt3\over3}$ 
                       & 0 & 0 & ${2\sqrt2\over3}$ & $-{2\over3}$
                       & $-{4\over3}$ 
                       & $-{2\over3}$ \cr
\hline
\end{tabular}}  
\end{table}

\begin{table}
\caption{\label{tab:gun}
Reduced $B\to MS^*$ matrix elements for non-strange $s$-wave resonances}
{\renewcommand{\arraystretch}{1.8} 
\setlength{\tabcolsep}{5pt}        
\begin{tabular}{lrrrrrrrr} 
\hline
 $S^*/BM$ & $\Sigma\bar{K}$ 
       & $ N\pi$  
       & $\Lambda\bar{K}$ 
       & $ N\eta_1$  
       & $ N\eta_8$ 
       &&\cr
\hline
$S11_2$  & $-{1\over3}\sqrt{2\over3}$ & $-{4\over3}$ & $-\sqrt2$ 
                & ${2\sqrt2\over3}$ & ${2\over3}$ &&\cr
$S11_4$ & $-{4\over3}\sqrt{2\over3}$ & $ {2\over3}$ & 0 
                & ${2\sqrt2\over3}$ & ${2\over3}$  &&\cr
$S31$           & ${2\over3}\sqrt{2\over3}$ & $ {2\over3}$ & 0 & 0 &&&\cr
\hline
\end{tabular}}  
\end{table}

\section{Contact interaction}

For the $s$-wave mesons the contact interaction can be cast in the form
\begin{eqnarray}
    V^c_{\alpha\beta}(k,k') &=&  {g_{\alpha\beta}\over2f^2_\pi} 
    {kk'\over 2\pi^2\sqrt{2\omega_\alpha(k)}\sqrt{2\omega_\beta(k')}}
\nonumber\\
   &&\kern-60pt\times      [\omega_\alpha(k) + \omega_\beta(k')]
    \int_0^R dr \,r^2[u_s^2(r) + v_s^2(r)]j_0(kr)j_0(k'r)\,.
\nonumber\\
\end{eqnarray}
Here $g_{\alpha\beta}$ can be identified with  $-2f_\pi^2\lambda^I_{\alpha\beta}$
of \cite{Thomas85} and $\half\mathcal{D}_{\alpha\beta}$ of 
\cite{Oset98} and are collected in table~\ref{tab:CI}.

\begin{table}[h]
\caption{\label{tab:CI} $g_{\alpha\beta}$ for isospin 0 and 1.}
$$
I=0
\qquad{\renewcommand{\arraystretch}{1.5} 
\setlength{\tabcolsep}{12pt}        
\begin{array}{|c|c|cccc|}
\hline
g_{\alpha\beta} &  KN &  \bar{K}N & \pi\Sigma  & \eta\Lambda & K\Xi \\
\hline
 KN          & 0 &&&&\\
\hline
 \bar{K}N    && {3\over2}     &{\sqrt6\over4}&{3\over\sqrt8} & 0 \\
 \pi\Sigma   && {\sqrt6\over4}& 2            & 0             &{\sqrt6\over4}\\
 \eta\Lambda && {3\over\sqrt8}& 0            & 0             &-{3\over\sqrt8}\\ 
  K\Xi       && 0             &{\sqrt6\over4}&-{3\over\sqrt8}& {3\over2}\\
\hline 
\end{array}}
$$

$$
I=1
\qquad{\renewcommand{\arraystretch}{1.4} 
\setlength{\tabcolsep}{12pt}        
\begin{array}{|c|c|ccccc|}
\hline
g_{\alpha\beta} & KN & \bar{K}N & \pi\Sigma & \pi\Lambda & K\Xi & \eta\Sigma\\
\hline
 KN         &-1& &&&& \\
\hline      
 \bar{K}N   && {1\over2} & {1\over2} & -{\sqrt6\over4} & 0 & -{\sqrt6\over4}\\
 \pi\Sigma  && {1\over2} &  1        & 0                & 1 & 0 \\
 \pi\Lambda && -{\sqrt6\over4} & 0   & 0 & -{\sqrt6\over4} &  0 \\
  K\Xi      &&  0        & 1 & -{\sqrt6\over4} & 1 &-{\sqrt6\over4} \\ 
\eta\Sigma  &&-{\sqrt6\over4}& 0 & 0 & -{\sqrt6\over4} & 0 \\ 
\hline 
\end{array}}
$$
\end{table}
Oset \cite{Oset98} has an opposite sign for 
$\bar{K}N\leftrightarrow\pi\Sigma$, which is compensated by changing 
the sign for $\Sigma^*\to\pi\Sigma$.

\end{document}